\documentclass[%
 reprint,
 amsmath,amssymb,
 aps,
prl,
]{revtex4-2}

\usepackage{graphicx}% Include figure files
\usepackage{dcolumn}% Align table columns on decimal point
\usepackage{bm}% bold math
\usepackage{braket}
\usepackage{xcolor}
\begin{document}

\preprint{APS/123-QED}

\title{Hyperparametric Oscillation via Bound States in the Continuum}% Force line breaks with \\

\author{Fuchuan Lei}
 \affiliation{Department of Microtechnology and Nanoscience, Chalmers University of Technology SE-41296 Gothenburg, Sweden }%Lines break automatically or can be forced with \\

\author{Zhichao Ye}
 \affiliation{Department of Microtechnology and Nanoscience, Chalmers University of Technology SE-41296 Gothenburg, Sweden }%Lines break automatically or can be forced with \\

\author{Krishna Twayana}%
\affiliation{Department of Microtechnology and Nanoscience, Chalmers University of Technology SE-41296 Gothenburg, Sweden
}%

\author{Yan Gao}%
\affiliation{Department of Microtechnology and Nanoscience, Chalmers University of Technology SE-41296 Gothenburg, Sweden
}%

\author{Marcello Girardi}%
\affiliation{Department of Microtechnology and Nanoscience, Chalmers University of Technology SE-41296 Gothenburg, Sweden
}%

\author{Óskar B. Helgason}%
\affiliation{Department of Microtechnology and Nanoscience, Chalmers University of Technology SE-41296 Gothenburg, Sweden
}%

\author{Ping Zhao}%
\affiliation{Department of Microtechnology and Nanoscience, Chalmers University of Technology SE-41296 Gothenburg, Sweden
}%

%\author{Ping Zhao}%
%\affiliation{Department of Microtechnology and Nanoscience, Chalmers University of Technology SE-41296 Gothenburg, Sweden
%}%

\author{Victor Torres-Company}%
 \email{torresv@chalmers.se}
\affiliation{Department of Microtechnology and Nanoscience, Chalmers University of Technology SE-41296 Gothenburg, Sweden
}%

\date{\today}% It is always \today, today,

\begin{abstract}
Optical hyperparametric oscillation based on the third-order nonlinearity is one of the most significant mechanisms to generate coherent electromagnetic radiation and produce quantum states of light. Advances in dispersion-engineered high-$Q$ microresonators allow for generating signal waves far from the pump and decrease the oscillation power threshold to submilliwatt levels. However, the pump-to-signal conversion efficiency and absolute signal power are low, fundamentally limited by parasitic mode competition and  attainable cavity intrinsic $Q$  to coupling $Q$ ratio, i.e., $Q_{\rm i}/Q_{\rm c}$. Here, we use Friedrich-Wintgen bound states in the continuum (BICs) to overcome the physical challenges in an integrated microresonator-waveguide system. As a result, on-chip coherent hyperparametric oscillation is generated in BICs with unprecedented conversion efficiency and absolute signal power. This work not only opens a path to generate high-power and efficient continuous-wave electromagnetic radiation in Kerr nonlinear media but also enhances the understanding of microresonator-waveguide system - an elementary unit of modern photonics.
\end{abstract}

\maketitle
\begin{figure}[b]
\centering
\includegraphics[width=1\linewidth]{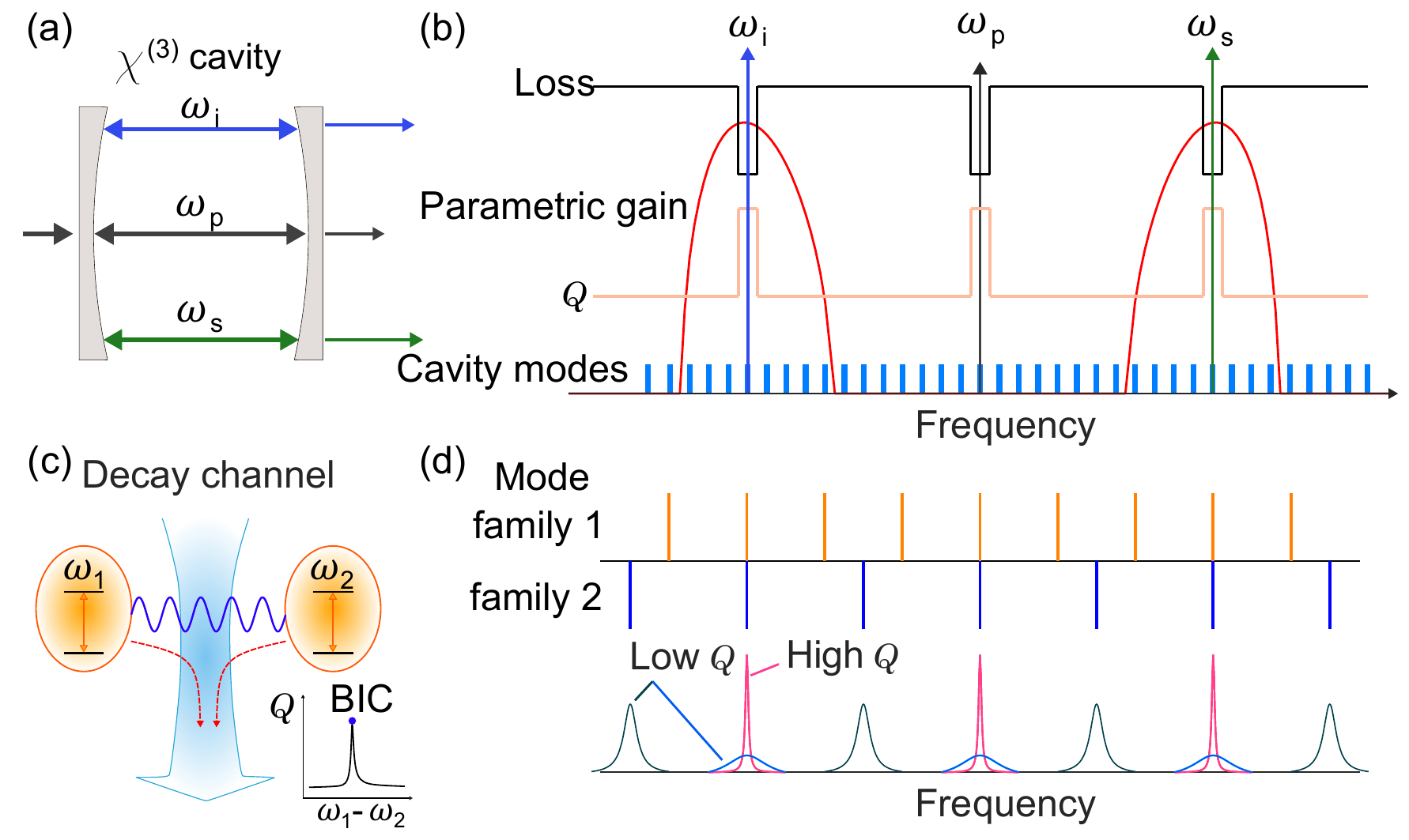}% Here is how to import EPS art
\caption{Concept of  H-OPO via bound states in the continuum (BICs). (a) Schematic of H-OPO. (b) Cavity quality factor ($Q$) management for achieving high-efficiency H-OPO.  (c) Schematic illustration of the Friedrich-Wintgen BIC. Two near degenerate resonances can be dissipatively coupled if both share a common decay channel. As a result, two supermodes are formed: one is a nondecaying bound state (high $Q$) while the other one is an increased-decaying state (low $Q$). (d) Realization of ($Q$) engineering for cavity modes by making  use of BICs in a two-mode cavity, where the high-$Q$ mode can be achieved periodically owing to the Vernier effect.}
\label{fig1}
\end{figure}

Optical hyperparametric oscillation (H-OPO) emerges in a driven $\chi^{(3)}$
nonlinear cavity as a result of modulation instability (MI) that amplifies vacuum photons. Two pump photons are converted into a  pair of correlated photon pairs ($2\hbar\omega_{p} \rightarrow \hbar\omega_s+\hbar\omega_i$) at new frequencies (Fig. \ref{fig1}(a)). This elementary mechanism lies at the onset of microresonator frequency comb generation \cite{del2007optical} and has served for generating coherent light sources \cite{kippenberg2004kerr,savchenkov2004low,wang2013mid,yang2016four,shen2018low,tian2022blue} and quantum technologies \cite{dutt2015chip,chembo2016quantum,kues2017chip,imany201850,matsko2019hyperparametric,lu2019chipNP,okawachi2020demonstration}. High-$Q$ microresonators, either in whispering gallery mode \cite{lin2017nonlinear,li2018whispering}, planar \cite{levy2010cmos,razzari2010cmos,hausmann2014diamond,pu2016efficient,li2016efficient,ji2017ultra,tang2020widely,wang2021high} or photonic crystal cavities \cite{marty2021photonic}, provide means to decrease the power oscillation threshold. With dispersion engineering, broadband  H-OPO has been demonstrated \cite{sayson2019octave,fujii2019octave,lu2019milliwatt,lu2020chip,domeneguetti2021parametric}.

\begin{figure*}[t]
\centering
\includegraphics[width=0.85\linewidth]{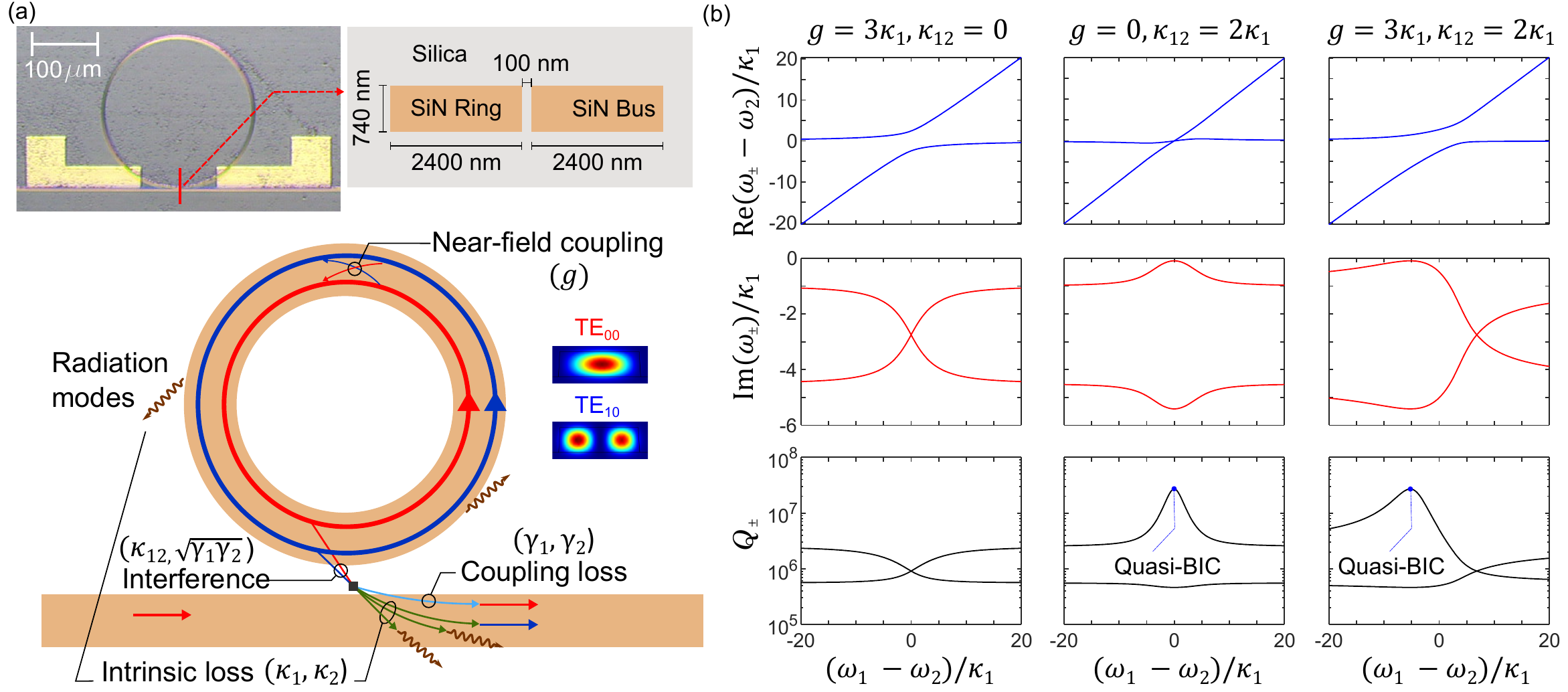}% Here is how to import EPS art
\caption{BICs in a multimode microring-waveguide system. (a) Image and schematic of the system. (b) Calculated eigenfrequencies of $\mathcal{H}$ and the corresponding $Q$ as a function of the detuning. Parameters used here: $\gamma_1=\gamma_2=0$, $\kappa_2=4.5\kappa_1$, $\omega_1/\kappa_1=5\times10^6$. }
\label{fig2}
\end{figure*}

A critical metric in any parametric oscillator is the conversion efficiency, i.e., the ratio of signal power compared to the pump. This metric is first limited by mode competition among different longitudinal modes in the cavity and nondegenerate four-wave mixing \cite{stone2022conversion}. The gain required to sustain H-OPO originates from MI. For a fixed dispersion, the gain's peak and bandwidth are determined by the pump power \cite{radic2008parametric,ye2021overcoming}. With high pump power, the gain can span multiple free spectral ranges (FSRs), even for extremely small cavities \cite{herr2012universal,victor2014comparative}, as schematically illustrated in Fig. 1(b).  Hence, multiple cavity modes could emit simultaneously once their losses are compensated by the gain. The newly generated frequency components could induce by themselves parametric oscillation, and degrade the coherence through nondegenerate four-wave mixing  \cite{herr2012universal}.  To avoid these detrimental effects, one could engineer the resonator's dispersion and pump it with moderate power \cite{sayson2019octave,kuo2012dispersion,fujii2019octave,lu2019milliwatt,lu2020chip,domeneguetti2021parametric}. Another strategy is to reduce the number of effective cavity modes, such as increasing the FSR of microresonators  \cite{lu2019milliwatt,stone2022conversion} or suppressing undesired modes with the aid of frequency selection elements (see Fig. 1(b)), e.g., coupled cavities and gratings \cite{gentry2014tunable,lu2022high}. These classical techniques, being widely used for single-frequency lasers, however, cannot be directly adapted for nonlinear optics because they are usually incompatible with the stringent demand of  simultaneously attaining high-$Q$, low-mode volume and dispersion engineering. 
In addition to multimode interaction, the second limiting factor of the conversion efficiency is the attainable 
 $Q_{\rm i}/Q_{\rm c}$ \cite{sayson2019octave}, where $Q_{\rm i(c)}$ is the cavity intrinsic (coupling) quality factor. Realistic microresonator-coupler systems introduce a physical limit because the coupling rate ($\sim 1/Q_{\rm c}$) cannot be arbitrarily high without involving additional intrinsic loss, hence, decreasing $Q_{\rm i}$ \cite{spencer2014integrated, pfeiffer2017coupling}.

%such as shrinking the size of Considering the FSR of microresonators cannot be made arbitrary large, another strategy is to suppress the unwanted modes and maintain desired modes by mode selection elements, see Fig. 1(b).

In this Letter, we demonstrate that the abovementioned limits can be overcome simultaneously by exploiting the concept of bound states in the continuum (BICs). As a result,  high-efficiency  H-OPO  could be accomplished. BICs were originally proposed by von Neumann and Wigner nearly one century ago in quantum physics and in recent years they have been extended to many other fields \cite{hsu2016bound,azzam2021photonic}. In photonics, BICs can be used to trap light and obtain high-$Q$ modes \cite{lepetit2010resonance,gentry2014dark,zou2015guiding,rybin2017high,bezus2018bound,yu2019photonic}, leading to applications such as lasing \cite{hodaei2016dark,kodigala2017lasing,huang2020ultrafast}, sensing \cite{yanik2011seeing,zhen2013enabling} and nonlinear optics \cite{carletti2018giant,koshelev2020subwavelength,zakharov2020transverse,liu2019high,krasikov2018nonlinear,wang2017improved}. Here we implement the Friedrich-Wintgen BICs in an integrated multimode microresonator-waveguide system. As schematically shown in Fig. \ref{fig1} (c), Friedrich-Wintgen BICs can occur as a result of destructive interference if two near degenerate resonances are dissipatively coupled through a common decay channel \cite{friedrich1985interfering,dittes2000decay}. Mathematically, Friedrich-Wintgen BICs are one type of singularity in anti-parity-time symmetric systems \cite{hodaei2016dark,miri2019exceptional,peng2016anti,jiang2019anti,choi2018observation, yang2020unconventional,li2019anti}. To utilize the concept of BICs for the mode number-dependent $Q$ factor management,  we take advantage of the Vernier effect in multimode microresonators (Fig. \ref{fig1}(d)).

\begin{figure*}[ht]
\centering
\includegraphics[width=0.85\linewidth]{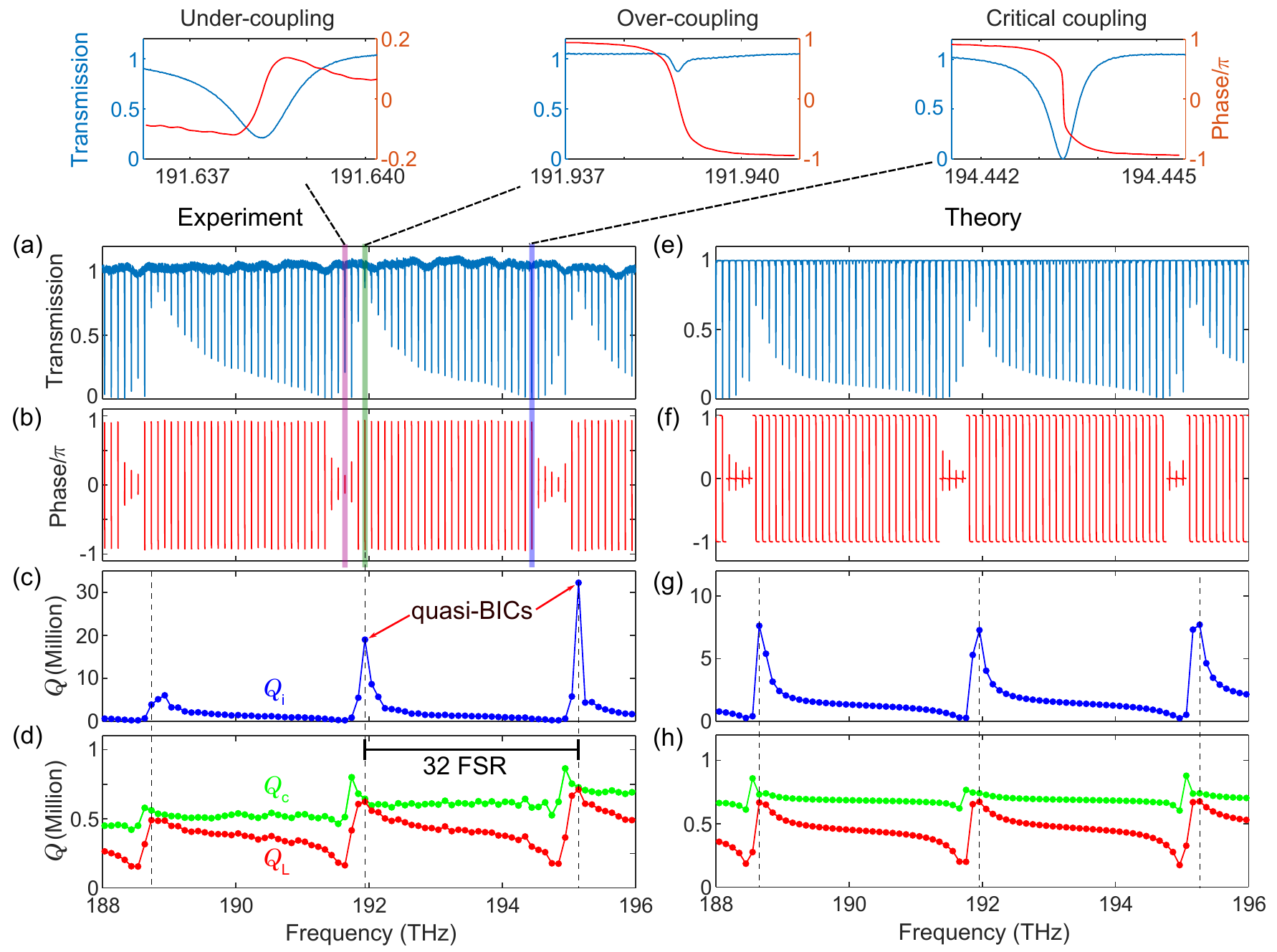}% Here is how to import EPS art
\caption{ Observation of BICs via spectral characterization of the system. (a) Normalized transmission scan of the microring-waveguide system. (b) Effective phase response of the system. (c) Intrinsic and (d) coupling as well as loaded $Q$ factors for the measured resonances. Three coupling conditions: Undercoupling, overcoupling and critical coupling are enlarged in the top panel. (e)-(h) Theoretical results corresponding to the case in panels (a)-(d).}
\label{fig3}
\end{figure*}
 
We consider an experimental system which consists of a silicon nitride ($\rm Si_3N_4$) microring  resonator and an adjacent bus waveguide, as shown in Fig. \ref{fig2}(a). Both the ring and bus waveguide support the higher-order modes $\rm TE_{10}$ and $\rm TE_{20}$  in addition to the fundamental mode $\rm TE_{00}$. The $\rm TE_{20}$ can be ignored in practice because its $Q$ is much lower. Unlike most widely used microring-waveguide systems, the ring-bus gap here is very small, which allows the fundamental cavity modes to couple with the other (guided and radiation) modes in the bus waveguide besides the fundamental one, see Fig. \ref{fig2}(a). This results in parasitic loss for the cavity mode and an encompassing reduction in $Q_{\rm i}$ and coupling ideality \cite{pfeiffer2017coupling}. The role of the bus waveguide on the intrinsic property of microcavities is not sufficiently appreciated \cite{lei2020polarization}, but it can have a dramatic influence, especially in multimode cavities  \cite{li2019multimode,ji2021exploiting,zhang2022ultralow}. In particular, the parasitic loss caused by the bus waveguide can be coherently suppressed when two near resonant cavity modes exist. In this case, Friedrich-Wintgen BICs could emerge because two cavity modes are coupled with the same decay channels. The motion of two near resonant cavity modes can be described by a Schrödinger-type equation (see Supplement Material for more details on the theoretical model and measurements, which includes Refs. \cite{gardiner2004quantum,xiao2010asymmetric,ye2019high,twayana2021frequency})
\begin{equation}
    i\frac{\partial}{\partial t}\ket{\psi}= \mathcal{H}\ket{\psi}+\ket{s},
\end{equation}
\begin{equation}
   \mathcal{H}=\begin{pmatrix}\omega_1 &g\\g^*&\omega_2
     \end{pmatrix}
     -i\begin{pmatrix}\kappa_1 &\kappa_{12}\\\kappa^*_{12}&\kappa_2 
     \end{pmatrix}
     -i\begin{pmatrix}\gamma_1 &\sqrt{\gamma_1\gamma_2}\\\sqrt{\gamma_1\gamma_2}&\gamma_2
     \end{pmatrix}
\end{equation}
with $\ket{\psi}=[a_1,a_2]^T$, $\ket{s}=[\sqrt{2\gamma_1},\sqrt{2\gamma_2}]^Ts_{\rm in}$, where the $a_{1(2)}$ are the complex amplitudes of the $\rm TE_{00}$ and $\rm TE_{10}$ cavity modes. In the first term of  $\mathcal{H}$, $\omega_{1(2)}$ are the resonant frequencies in the uncoupled system, and $g$ is the scattering-induced direct coupling coefficient between the two modes. The second term of  $\mathcal{H}$ is non-Hermitian, where $\kappa_{1(2)}$ 
stand for the decay rates caused by intrinsic loss including material absorption, radiation loss and bus-waveguide-induced parasitic loss. $\kappa_{12}$ denotes  the via-the-continuum coupling term since the two cavity modes share the same decay channels. This is the critical parameter for achieving high-$Q_{\rm i}$ BIC modes. We note that the value of  $\kappa_{12}$ is restricted to $|\kappa_{12}|< \sqrt{\kappa_1\kappa_2}$ as not all decaying terms can be canceled by perfect destructive interference, i.e., only quasi-BICs can be attained. 
The third term of $\mathcal{H}$ describes the two cavity modes coupling to the fundamental mode of the bus waveguide at rates $\gamma_{1(2)}$. The term  $\sqrt{\gamma_1\gamma_2}$ plays a role similar to $\kappa_{12}$ to generate BICs but it acts on $Q_{\rm c}$ instead of $Q_{\rm i}$ \cite{gentry2014dark}. It is worth noting that $Q_{\rm i}$ and $Q_{\rm c}$ play the same role in $\mathcal{H}$, but their impact on the light coupling and thus the efficiency of the nonlinear optics process is quite different.

BICs can be obtained from the eigenvalues ($\omega_{\pm}$) of $\mathcal{H}$. The real and imaginary parts of the eigenvalues stand for the resonant frequencies and decay rates of two eigenmodes formed by the superposition of the original cavity modes.  The quality factors of the two eigenmodes can be calculated by $Q_{\pm}=|{\rm Re}(\omega_{\pm})/2{\rm Im}(\omega_{\pm})|$. Because of coupling, both of the resonant frequencies and decay rates vary with the detuning ($\omega_1-\omega_2$). To illustrate that $Q_{\rm i}$ can be tailored as a result of BICs, we assume $\gamma_{1(2)}=0$ and plot the complex frequencies as a function of the detuning, see Fig. 2(b). 
It is shown that the scattering-induced near field direct coupling ($g$) and via-the-continuum indirect coupling ($\kappa_{12}$) lead to distinct effects on the eigenvalues. 
The former leads to the real parts of eigenvalues being avoided and imaginary parts crossed, while the latter gives rise to the opposite effect. Moreover, when only one eigenvalue becomes near purely real, a quasi-BIC is formed at the expense of the other eigenvalue becoming more lossy (its imaginary part increases). In a realistic system, both direct coupling and indirect coupling exist, thus the interference at the coupling region is 
determined by both. In this scenario, quasi-BICs can still be obtained but, surprisingly at a nonzero detuning, while the $Q$s also exhibit asymmetric dependence on the detuning. These aspects will be demonstrated in the following experiment.

The bus waveguide features a tapered structure at both ends to ensure only $\rm TE_{00}$ can be excited and collected. The measured normalized transmission spectrum of the system is shown in Fig. \ref{fig3}(a). Unlike the conventional transmission spectrum of a microring-waveguide system, a doubly periodic pattern shows up, given by the nominal FSR of the fundamental mode and a periodic deletion the transmission resonance that arises from the interaction between between $\rm TE_{00}$ and $\rm TE_{10}$ mode families. The period is 32 FSRs of $\rm TE_{00}$ mode, which matches the Vernier frequency corresponding to the walk-off between group indices  (see Supplemental Material).

Although two transverse modes families are involved, it is noted that only a single clear resonance dip can be observed within each FSR  dominated by  $\rm TE_{00}$ family modes. This can be interpreted as the low-$Q_{\rm i}$ eigenmodes being extremely undercoupled. Therefore, the transmission spectrum and phase response can be approximately considered as the consequence of single high-$Q_{\rm i}$ eigenmodes. The measured transmittance and phase  allow us to distinguish unambiguously $Q_{\rm i}$ and $Q_{\rm c}$, see Fig. \ref{fig3} and Figure S2 in the Supplemental Material  \cite{twayana2021frequency}.

Following the transmission spectrum, both  $Q_{\rm i}$ and $Q_{\rm c}$ feature periodic patterns but with slightly shifted frequency dependence, which could be explained as the intrinsic loss and coupling loss of cavity modes correspond to different decay channels. As discussed above, due to the coexistence of direct and indirect coupling, both $Q_{\rm i}$ and $Q_{\rm c}$ as well as the transmission spectrum exhibit a Fano-like frequency dependence. The  
asymmetric $Q$ distribution with respect to pump mode further reduces the risk of mode competition in H-OPO as demonstrated in the following section. In addition to suppressing mode competition, another advantage of engineering $Q_{\rm i}$ is that the modes with highest $Q_{\rm L}$ are strongly overcoupled (top panel of Fig. 3). This counterintuitive phenomenon is especially important for achieving high-efficiency nonlinear optics phenomena, and H-OPO in particular.

The above theoretical model only considers the interaction between a pair of near resonant modes. To better describe the realistic system over a wider spectral range, we have generalized the above  model to four interacting modes (Supplemental Material) by including
the effect that one mode from $\rm TE_{00}$ family could couple to more than one mode from the $\rm TE_{10}$ family simultaneously and vice versa, see Figs. 3(e)-(h). The good match between theory and experiment suggests our model captures the main underlying physics of the system.

\begin{figure}[t] 
\centering
\includegraphics[width=0.95\linewidth]{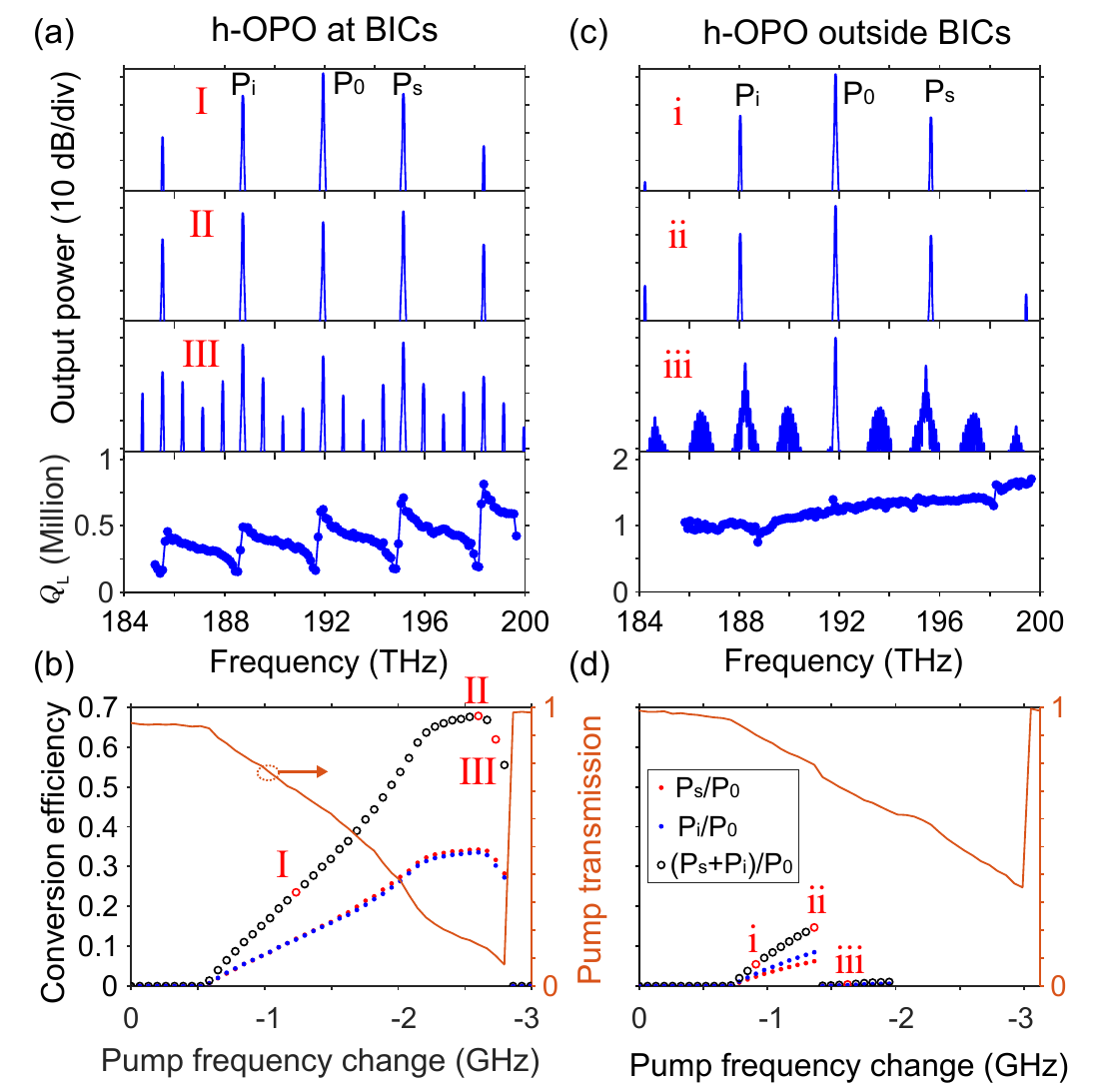}% Here is how to import EPS art
\caption{\label{Fig4_7} {Experimental demonstration of H-OPO in microresonators at and outside BICs. (a) Measured optical spectra  at different pump-cavity detunings. As a reference, frequency-dependent loaded $Q$ ($Q_{\rm L}$) is plotted at the bottom. (b) Calculated conversion efficiency and pump transmission as a function of the pump frequency. (c)-(d) same as (a)-(b) except for pumping another sample.}} 
\end{figure}

To generate H-OPO, anomalous dispersion is required. We found the supermode's dispersion is dominated by the $\rm TE_{00}$ mode with mode-interaction-induced distortion (see Supplemental Material). To pump this device, we use a laser located at 1561.9 nm after being amplified by an erbium-doped fiber amplifier. The on-chip pump power threshold of this H-OPO is $\sim$60 mW, and the highest conversion efficiency is attained at 200 mW. To couple pump into resonance, we decrease laser's frequency gradually. With more pump power coupled into the microresonator, a pair of signal and idler waves are generated, exactly located at the quasi-BIC locations, see Fig. 4(a). With further decreasing the pump frequency, the power of signal and idler increases linearly before reaching saturation (see Fig. 4(b)). The maximum conversion efficiency $\rm \eta_s=(P_{s}+P_{i})/P_{0}=68\%$, corresponding to $\sim$ 70 mW on-chip signal power. The value could be further improved if the undesired BICs (i.e. BICs modes excluding pump, signal and idler modes) and the resultant cascaded four-wave mixing were suppressed (Supplemental Material), for example, by coupling with an auxiliary cavity \cite{zhang2021squeezed}. By further reducing the detuning, the conversion efficiency decreases suddenly as a result of the birth of new lines, indicating the low-$Q$ modes can still be activated at high pump power. The 
advantage of BICs can be clearly shown if we pump another sample, which has the same geometric parameters except for 50 nm larger ring-bus gap. Hence, it does not feature quasiperiod $Q$ response and the highest conversion efficiency is limited to $ <15\%$ due to mode competition.

To conclude, we have implemented BICs in an integrated Kerr nonlinear microring resonator featuring a high-contrast quasiperiodic response in quality factor. By exploring waveguide-assisted dissipative mode coupling in   microring-waveguide system, we generalize the spectral response and demonstrate the feasibility of boosting $Q_{\rm i}$ in high-$Q$ microresonators. Such a simple system not only enables the unwanted modes to be suppressed, but it also allows to obtain strong overcoupling without sacrificing $Q_{\rm i}$ for desired modes. The $Q_{\rm i}$ can in fact be higher than in standard single-mode systems, resulting in unprecedented efficiency in H-OPO and high-power signal and idler waves. Besides H-OPO, the possibility to engineer the spectral response of the intrinsic quality factor of high-Q microresonators can be useful for other power-efficient nonlinear optics scenarios, and the strong overcoupling (high $Q_{\rm i}/Q_{\rm c}$) could especially facilitate the generation of squeezed or entangled states of light \cite{dutt2015chip,zhao2020near,yang2021squeezed,zhang2021squeezed,sabattoli2021suppression}.

The devices demonstrated in this work were
fabricated at Myfab Chalmers. This work was supported by European Research Council (CoG GA 771410); Vetenskapsrådet (2020-00453). The raw data for this work can be accessed at https://doi.org/10.5281/zenodo.7529404.  

\newpage
\newpage
\section{Theory}
\subsection{Two-mode model}
The Hamiltonian of a two-mode (one $\rm TE_{00}$ and one $\rm TE_{10}$) microresonator  coupled with a bus waveguide system reads $(\hbar=1)$ 
\begin{equation*}
\begin{split}
    \textbf{H}=&\sum_{k=1,2}\omega_k a_k^\dag a_k+
    (ga^\dag_1 a_2+g^*a^\dag_2 a_1)  +
    \int_{-\infty}^{\infty}  \omega b^\dag(\omega) b(\omega) d\omega \\
& +\sum_j\int_{-\infty}^{\infty} \omega c^\dag_j(\omega) c_j(\omega) d\omega \\
    &+\sum_{k=1,2}\int_{-\infty}^{\infty} d\omega [\eta_k(\omega)a_k^\dag b(\omega)+ \eta_k^*(\omega)a_kb^\dag(\omega)] \\
    &+\sum_{k=1,2}\sum_{j}\int_{-\infty}^{\infty} d\omega [\lambda_{kj}(\omega)a_k^\dag c_j(\omega)+ \lambda_{kj}^*(\omega)a_kc^\dag_j(\omega)],
    \label{700}
\end{split}
\end{equation*}
where $a_{k} [a^\dag_{k}]$ is the annihination (creation) operator of the $k$th cavity mode, whose resonant frequency is $\omega_k$. $g$ is the
scattering-induced (e.g., roughness induced scattering) direct coupling coefficient between  the two cavity modes. $b(\omega)  [b^\dag(\omega)]$ and $c_j(\omega)[c_j^\dag(\omega)]$ denote the annihilation (creation) operator of the fundamental mode of bus waveguide ($\rm TE_{00}$) and $j$th other modes (all guided modes of the bus waveguide except for $\rm TE_{00}$ and radiation modes), which are coupled with $k$th cavity mode with coupling coefficients $\eta_k$ and $\lambda_{kj}$ respectively. 

The Heisenberg equations of motion for the cavity and waveguide modes are
\begin{equation}
\begin{split}
    \frac{da_1}{dt}=&-i\omega_1a_1-iga_2-i\int_{-\infty}^{\infty} d\omega \eta_{1}(\omega)b(\omega)   \\
&-i\sum_j\int_{-\infty}^{\infty} d\omega \lambda_{1,j}(\omega)c_j(\omega),
      \label{eq701a}
\end{split}
\tag{S.1}
\end{equation}

\begin{equation}
\begin{split}
    \frac{da_2}{dt}=&-i\omega_2a_2-ig^*a_1-i\int_{-\infty}^{\infty} d\omega \eta_{2}(\omega)b(\omega) \\
  &  -i\sum_j\int_{-\infty}^{\infty} d\omega \lambda_{2,j}(\omega)c_j(\omega),
      \label{eq701b}
      \end{split}
      \tag{S.2}
\end{equation}

\begin{equation}
    \frac{db(\omega)}{dt}=-i\omega b(\omega)- i\sum_{k=1,2} \eta^*_{k}(\omega)a_k,
    \label{eq702}
    \tag{S.3}
\end{equation}
\begin{equation}
    \frac{dc_j(\omega)}{dt}=-i\omega c_j(\omega)-  i\sum_{k=1,2} \lambda^*_{k,j}(\omega)a_k.
    \label{eq703}
    \tag{S.4}
\end{equation}
By integrating Eqs. (\ref{eq702}-\ref{eq703}), we obtain 

\begin{equation}
\begin{split}
b(\omega)=&b^0(\omega)e^{-i\omega (t-t_0)}-i\sum_{k=1,2}\eta_k^*(\omega) \int_{t_0}^t e^{-i\omega (t-t')}a_k(t')dt',
\label{eq704}
\end{split}
\tag{S.5}
\end{equation}

\begin{equation}
\begin{split}
c_j(\omega)=&c^0_j(\omega)e^{-i\omega (t-t_0)} -i\sum_{k=1,2}\lambda^*_{k,j}(\omega)\int_{t_0}^t e^{-i\omega (t-t')}a_k(t')dt',
\label{eq705}
\end{split}
\tag{S.6}
\end{equation}
 where $b^0(\omega)$ and $c^0_j(\omega)$ are the values of $b(\omega)$ and $c_j(\omega)$ at $t=t_0$. Bring Eqs. (\ref{eq704})-(\ref{eq705}) into Eqs. (\ref{eq701a})-(\ref{eq701b}), and we have

\begin{equation}
\begin{split}
     &\frac{da_1}{dt}=-i\omega_1a_1-iga_2-i\int_{-\infty}^{\infty} d\omega \eta_{1}(\omega)e^{-i\omega (t-t_0)}b^0(\omega) \\-&\int_{-\infty}^{\infty} d\omega \eta_{1}(\omega)\sum_{k=1,2}\eta_k^*(\omega) \int_{t_0}^t e^{-i\omega (t-t')}a_k(t')dt'\\
    -& i\sum_j\int_{-\infty}^{\infty} d\omega \lambda_{1,j}(\omega)e^{-i\omega (t-t_0)}c^0_j(\omega)\\
    -&\sum_j\int_{-\infty}^{\infty} d\omega \lambda_{1,j}(\omega)\sum_{k=1,2}\lambda^*_{k,j}(\omega)\int_{t_0}^t e^{-i\omega (t-t')}a_k(t')dt', 
          \label{eq706a}
\end{split}
\tag{S.7}
\end{equation}
\begin{equation}
\begin{split}
     &\frac{da_2}{dt}=-i\omega_2a_2-ig^*a_1-i\int_{-\infty}^{\infty} d\omega \eta_{2}(\omega)e^{-i\omega (t-t_0)}b^0(\omega)\\
     - & \int_{-\infty}^{\infty} d\omega \eta_{2}(\omega)\sum_{k=1,2}\eta_k^*(\omega) \int_{t_0}^t e^{-i\omega (t-t')}a_k(t')dt'\\
    - & i\sum_j\int_{-\infty}^{\infty} d\omega \lambda_{2,j}(\omega)e^{-i\omega (t-t_0)}c^0_j(\omega) \\- &\sum_j\int_{-\infty}^{\infty} d\omega \lambda_{2,j}(\omega)\sum_{k=1,2}\lambda^*_{k,j}(\omega)\int_{t_0}^t e^{-i\omega (t-t')}a_k(t')dt'. 
          \label{eq706b}
\end{split}
\tag{S.8}
\end{equation}
Using first-Markov approximation \cite{gardiner2004quantum,xiao2010asymmetric}, i.e., assuming $\eta_k$ and $\lambda_{k,j}$ to be constant as a function of $\omega$ (for convenience, $\eta_k$ can be assumed to be real by properly defining the phase of $a_k$),  
Eqs. (\ref{eq706a})-(\ref{eq706b}) can be reduced as
\begin{equation}
\begin{split}
       & \frac{da_1}{dt}=-i\omega_1a_1-iga_2-\gamma_{1}a_1-\sqrt{\gamma_1\gamma_2}a_2-i\sqrt{2\gamma_1}s_{\rm in} \\-&\kappa_1a_1-\kappa_{12}a_2 -i c_{1,{\rm in}},
       \end{split}
\label{eq707}
\tag{S.9}
\end{equation}
\begin{equation}
\begin{split}
    & \frac{da_2}{dt}=-i\omega_2a_2-ig^*a_2-\gamma_{2}a_2-\sqrt{\gamma_1\gamma_2}a_1 --i\sqrt{2\gamma_2}s_{\rm in}\\-  & \kappa_2a_2-\kappa_{12}^*a_1 -i c_{2,{\rm in}},
    \end{split}
          \label{eq708}
\tag{S.10}
\end{equation}
where $\gamma_k=|\eta_k|^2/2$,  $\kappa_k=\sum_j|\lambda_{k,j}|^2/2$, $\kappa_{12}=\sum_j\lambda_{1,j}\lambda^*_{2,j}/2$,
$s_{\rm in}=\int_{-\infty}^{\infty} d\omega e^{-i\omega (t-t_0)}b^0(\omega)$, $c_{k,{\rm in}}=\sum_j\lambda_{k,j}\int_{-\infty}^{\infty} d\omega e^{-i\omega (t-t_0)}c^0_j(\omega)$. $s_{\rm in}$ and $c_{k,{\rm in}}$ stand for the input field of the $k$th cavity mode. In the classical case, $c_{k,{\rm in}}$ can be ignored because the input field is only applied to the $\rm TE_{00}$ of the bus waveguide. In addition, the material absorption and ring bending radiation induced cavity decay should be considered, which can be included into $\kappa_{1,2}$ since they are treated as the intrinsic decay rate. It is easy to see $|\kappa_{12}|< \sqrt{\kappa_1\kappa_2}$. Equations. (\ref{eq707}-\ref{eq708}) can be expressed in the compact form, given by Eqs (1)-(2). 

\subsection{Four-mode model}
The above two-mode model captures the main underlying physics of BICs. 
However, it is incapable of reconstructing the transmission spectrum with a span beyond one Vernier period (32 FSRs). Owing to the Vernier effect, one  $\rm TE_{00}$ cavity mode could be equally coupled with two neighbouring longitudinal $\rm TE_{10}$ modes when its resonance frequency is in the middle. To generate a transmission spectrum with a span of three Vernier periods, we generalize the two-mode model to four-modes, where one $\rm TE_{00}$ mode simultaneously couples with three neighboring longitudinal modes of $\rm TE_{10}$ family, as illustrated in Figure S1(a). To be specific, we 
render $j$th ($j=1,2,3..$ from left to right) $\rm TE_{00}$ mode couple with $j$th, $j+1$th and $j+2$th $\rm TE_{10}$ modes simultaneously, and set the 1st $\rm TE_{00}$ mode to be red-detuned from the 1st $\rm TE_{10}$ mode. Such arrangement warranties one $\rm TE_{00}$ mode could couple with at least one $\rm TE_{10}$ mode with detuning less than 0.5 FSR, thus the feature of the transmission response can be well captured.

\begin{figure*}[!h]
\centering
\includegraphics[width=0.9\linewidth]{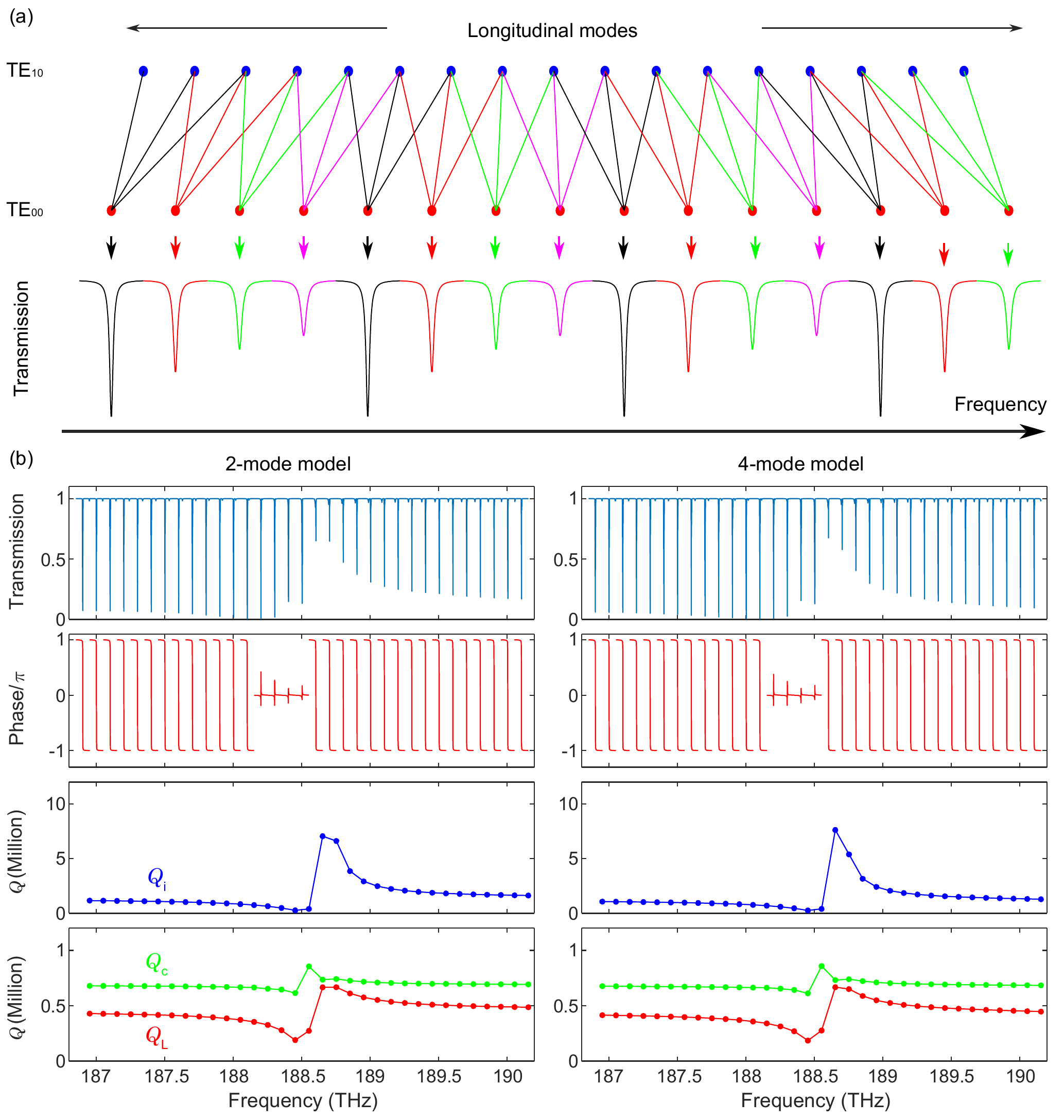}% Here is how to import EPS art
\caption{ Illustration of theoretical model. (a) Schematic of mode interaction in four-mode model. (b) Comparison of two-mode model and four-mode model. }
\label{fig:s1}
\end{figure*}

The coupled mode equations for the four-mode model can be directly generalized from Eqs. (1)-(2) as
\begin{widetext}
\begin{equation}
i\frac{d}{dt}\begin{pmatrix}
a\\
b_1\\
b_2\\
b_3
\end{pmatrix}
=\mathcal{H}\begin{pmatrix}
a\\
b_1\\
b_2\\
b_3
\end{pmatrix}+\begin{pmatrix}\sqrt{2\gamma_a}\\\sqrt{2\gamma_b}\\\sqrt{2\gamma_b}\\\sqrt{2\gamma_b}
\end{pmatrix}s_{\rm in},
\tag{S.11}
\end{equation}
\begin{equation*}
\begin{split}
  & \mathcal{H}= \begin{pmatrix}
\omega_a & g& g& g\\
g^*& \omega_{b1} &0 &0\\
g^*&0&\omega_{b2} &0\\
g^*&0&0 &\omega_{b3}
\end{pmatrix}-i\begin{pmatrix}
\kappa_a+\gamma_a & \kappa_{ab}+\sqrt{\gamma_a\gamma_b} & \kappa_{ab}+\sqrt{\gamma_a\gamma_b} &\kappa_{ab}+ \sqrt{\gamma_a\gamma_b} \\
\kappa^*_{ab}+\sqrt{\gamma_a\gamma_b} & \kappa_b+\gamma_b& \kappa_{bb}+\gamma_b & \kappa_{bb}+\gamma_b\\
\kappa^*_{ab}+\sqrt{\gamma_a\gamma_b} & \kappa^*_{bb}+\gamma_b& \kappa_b+\gamma_b& \kappa_{bb}+\gamma_b\\
 \kappa^*_{ab}+\sqrt{\gamma_a\gamma_b}& \kappa^*_{bb}+\gamma_b & \kappa^*_{bb}+\gamma_b & \kappa_b+\gamma_b\\
\end{pmatrix}
\end{split},
\tag{S.12}
\end{equation*}
\end{widetext}
where $a$ stands for the $\rm TE_{00}$ mode and $b_{1,2,3}$ 
correspond to three $\rm TE_{10}$  adjacent longitudinal modes. The coefficients $\omega$, $g$, $\gamma$ and $\kappa$ take on the similar physical meaning in the two-mode model.
 
The transmission spectrum around each $\rm TE_{00}$ resonances can be obtained 
$T=\left|{s_{\rm out}}/{s_{\rm in}}\right|^2$ with $s_{\rm out}$ being determined from cavity input-output relation \cite{gardiner2004quantum}
\begin{equation*}
    s_{\rm out}=s_{\rm in}-i\sqrt{2\gamma_a}a-i\sqrt{2\gamma_b}b_1-i\sqrt{2\gamma_b}b_2-i\sqrt{2\gamma_b}b_3.
    \tag{S.13}
\end{equation*}
To obtain the $Q_{\rm i}$ and $Q_{\rm L}$ of the high-$Q$ supermode, we use a Lorentz function to fit the deepest dip in the transmission spectrum. The phase response can be directly calculated from the transmission coefficient ${s_{\rm out}}/{s_{\rm in}}$.
The broadband transmission spectrum is formed by combing each individual spectrum. 

The parameters used for Fig. 3 are $\gamma_a/2\pi=140$ MHz, $\gamma_b/2\pi=10$ MHz, 
$\kappa_a/2\pi=70$ MHz, $\kappa_b/2\pi=1.7$ GHz, $g/2\pi=-900$ MHz, $\kappa_{ab}/2\pi=320$ MHz, $\kappa_{bb}/2\pi=1.6$ GHz. 

We emphasize the four-mode model and two-mode model only give little difference within one Vernier period. In Figure \ref{fig:s1}(b), we show a comparison of the spectral response given by the two-mode model and four-mode mode. We can see both predict quasi-BICs but with slight difference in transmission spectra and quality factors.  
\section{Device fabrication and characterization}
\begin{figure*}[h]
\centering
\includegraphics[width=0.9\linewidth]{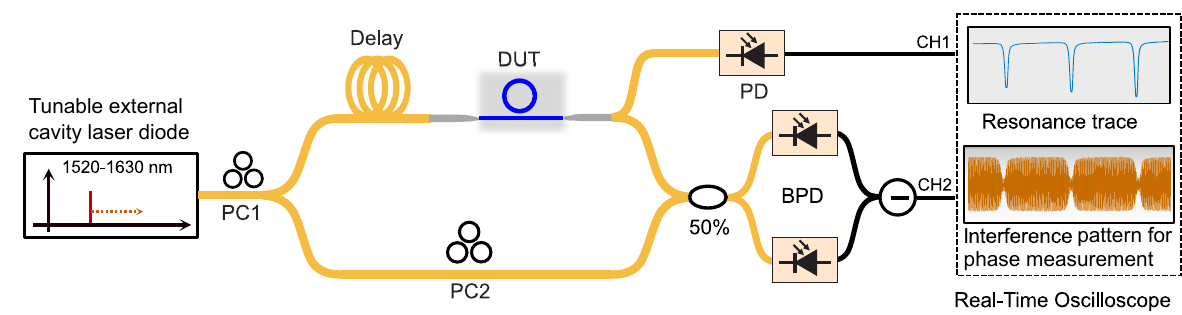}% Here is how to import EPS art
\caption{\label{fig:s2}  Setup for spectral characterization of microresonator-waveguide systems. PC: polarization controller; PD: photodiode; BPD: balanced photodiode. }
\end{figure*}
%\begin{figure}[h]
%\centering
%\includegraphics[width=1\linewidth]{couplingideality.pdf}% Here is how to import EPS art
%\caption{\label{fig:s3} Power transmission as a function of the linewidth
%(FWHM) of the modes. (a) The modes ranging from 185.2 nm to 188.5 THz of the sample used for Fig. 3. (b) The conventional case where the resonator-waveguide coupling strength varies.}
%\end{figure}

 The devices were fabricated on a silicon wafer using a subtractive processing method \cite{ye2019high}. The light is coupled into and out the chip using two lensed fibers with $5$ dB coupling loss in total. The devices were characterized by a tunable external cavity laser whose frequency is precisely measured by a frequency comb-calibrated fiber interferometer. To obtain the resonance prarameters and untangle undercoupling and overcoupling unambiguously, the complex transmission coefficient $t=\sqrt{T}e^{i\phi}$ was measured around each resonance, where $T$ is transmittance and $\phi$ is the phase. The phase is measured from swept-wavelength 
interference pattern, see Figure \ref{fig:s2} \cite{twayana2021frequency}. For each individual resonance, the measured $t$ as a function of laser frequency is fitted by 
\begin{equation}
t=1+\frac{\kappa_{\rm c}}{i(\omega-\omega_0)-\frac{\kappa_{\rm c}+\kappa_0}{2}+\frac{\kappa^2_{\rm i}}{i(\omega-\omega_0)-\frac{\kappa_{\rm c}+\kappa_0}{2}}},
 \tag{S.14}
\end{equation}
which is obtained from steady-state temporal coupled mode theory including back-scattering induced coupling (with coefficient $\kappa_{\rm i}$) between two degenerate counter-propagating modes. The evaluated coupling coefficients of the resonances enable
calculating the quality factors: $Q_{\rm i}=\omega_0/\kappa_{\rm i}$, $Q_{\rm c}=\omega_0/\kappa_{\rm c}$, and $Q_{\rm L}=\omega_0/(\kappa_{\rm i}+\kappa_{\rm c})$.

\begin{figure*}[t]
%\centering
%\includegraphics[width=0.8\linewidth]{Figs_TM.pdf}% Here is how to import EPS art
%\caption{\label{fig:s4}  Measured transmission spectrum of the sample used for Fig. 3 in the main text but with input polarization aligned to the TM modes. }
%\end{figure}
\centering
\includegraphics[width=0.8\linewidth]{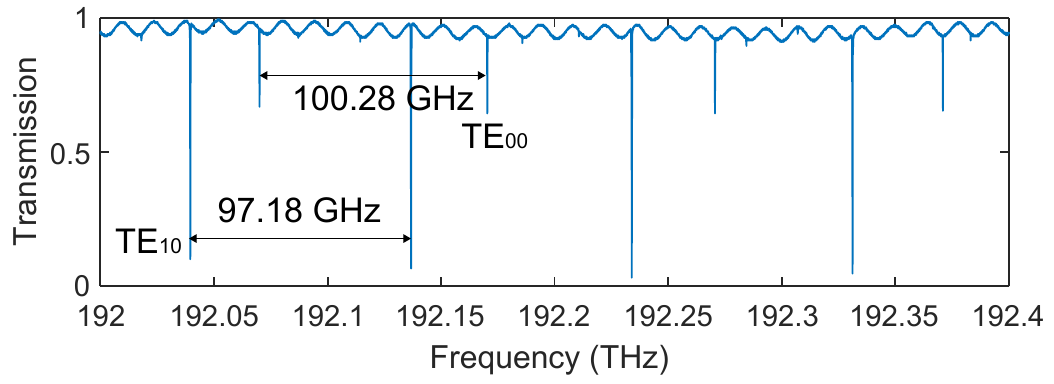}% Here is how to import EPS art
\caption{\label{fig:3} Measured transmission spectrum of a microring resonator coupled with a bus waveguide. The geometry of the microring is same as the one used for Fig. 3. The width of bus waveguide is 1200 nm and the bus-ring gap is 600 nm. The input polarization aligned to the TE modes. }
\end{figure*}

\section{Vernier effect in transmission spectrum}

 The observed double periodicity of transmission response in Fig. 3 can be attributed to the interaction between $\rm TE_{00}$ and $\rm TE_{10}$ mode families as the calculated Vernier period matches with the observation: ${f_{\rm V}/{\rm FSR(TE_{00})}}=n_{\rm g} ({\rm TE_{00}})/[n_{\rm g} ({\rm TE_{10}})-n_{\rm g} ({\rm TE_{00}})]=2.088/(2.154-2.088)=31.6$. Here $ n_{\rm g}$ is the group index simulated by COMSOL with measured material and geometric parameters. 

To obtain the FSRs of both $\rm TE_{00}$ and $\rm TE_{10}$ modes in the ring resonator experimentally, we measure another sample with the same dimension of ring  but different width of bus waveguide (1200 nm) and ring-bus gap (600 nm), as a result, both the two mode families can be directly observed, see Figure S3. The two families can be easily distinguished. The measured FSRs are 100.28 GHz and 97.18 GHz, which gives the Vernier period to be ${f_{\rm V}/{\rm FSR(TE_{00})}}={\rm FSR } ({\rm TE_{10}})/[{\rm FSR }({\rm TE_{00}})-{\rm FSR }({\rm TE_{10}})]=31.3$. Again, it matches with the pattern period in Fig. 3. 

\section{More information about microring-waveguide samples}

\begin{figure*}[h]
\centering
\includegraphics[width=0.9\linewidth]{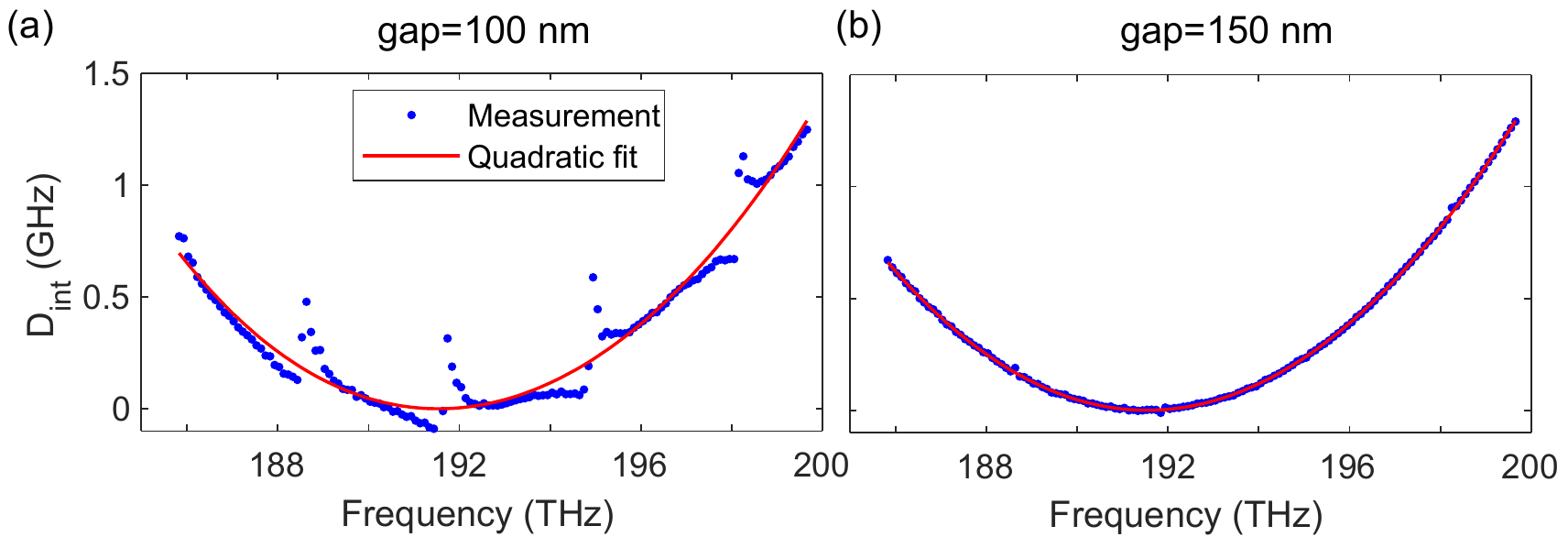}% Here is how to import EPS art
\caption{\label{fig:s4} Dispersion measurement. (a) Sample used for Fig. 3 in the main text. (b) Same size as (a) except for the microring-waveguide gap. }
\end{figure*}

\begin{figure*}[!h]
\centering
\includegraphics[width=1\linewidth]{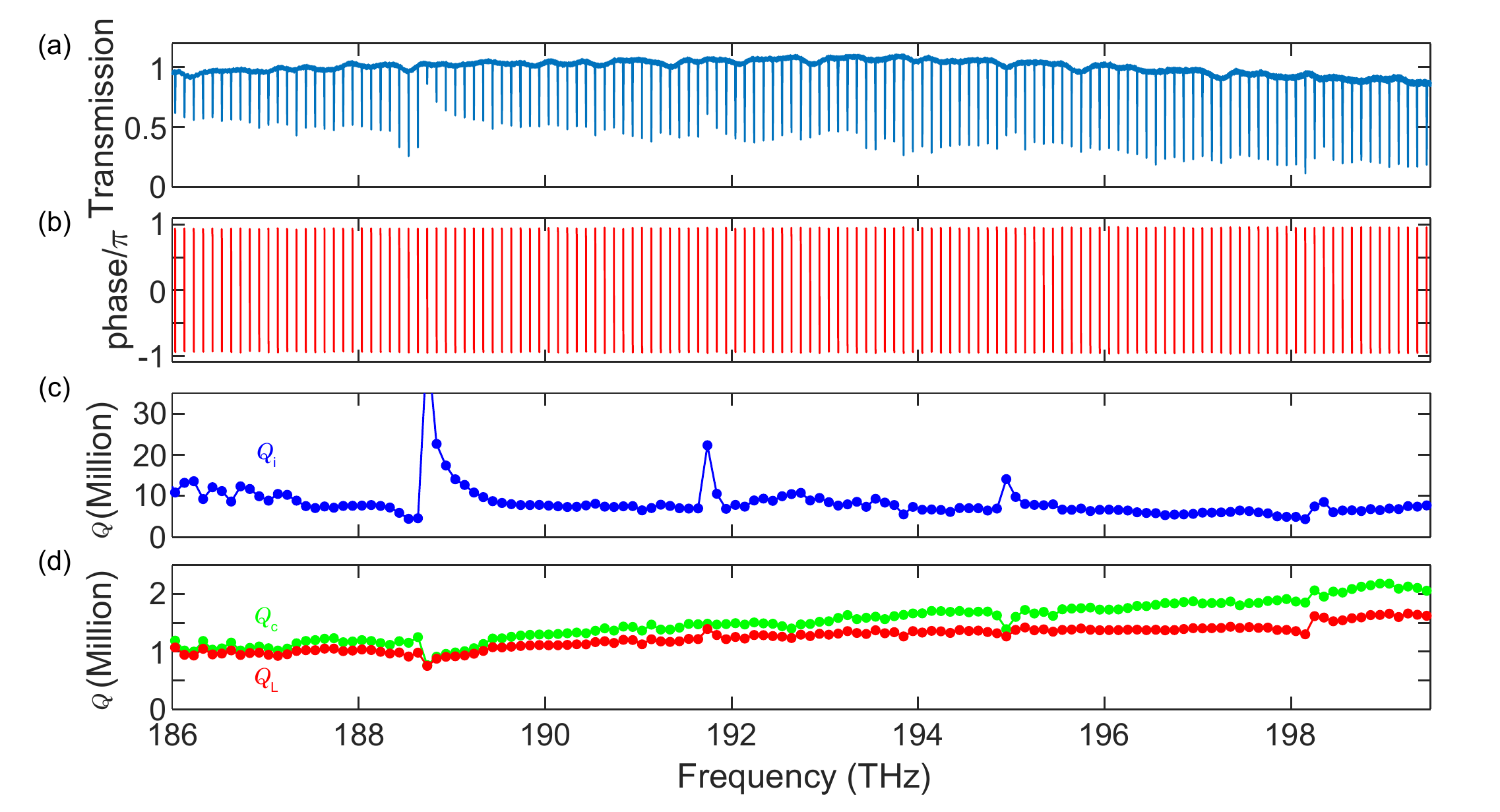}% Here is how to import EPS art
\caption{\label{fig:s6}Spectral characterization of the sample used for Fig. 4(c)-(d) in the main text. (a) Measured transmission spectrum (b) Phase respose (c) Intrinsic $Q_{\rm i}$ (d) Coupling $Q_{\rm c}$ and loaded $Q_{\rm L}$ quality factors. }
\end{figure*}

The dispersion of the cavity modes is characterized by the integrated dispersion $D_{\rm int}$:
\begin{equation}
    D_{\rm int}=\omega_{\mu}-(\omega_0+\mu D_1),
    \tag{S.15}
\end{equation}
where $\omega_{\mu}$ is the resonant frequency of $\mu$th longitudinal mode counting from the pump resonance. $D_1/2\pi $ is the microresonator FSR. By fitting $D_{\rm int}$ (Figure \ref{fig:s4}(a)), we obtained the group velocity dispersion $\rm \beta_2 =-46  ps^2/km$ for the sample used for Fig. 3 and Fig. 4(a)-(b) in the main text. For comparison, the dispersion of another sample used for Fig. 4(c)-(d) in the main text is  plotted in Figure \ref{fig:s4}(b). The two samples feature the same geometry except for microring-waveguide gap. By comparison, we can see $D_{\rm int}$ is dominated by $\rm TE_{00}$ mode family. This also suggests the measured high-$Q$ resonances are dominated by $\rm TE_{00}$ modes with perturbation by $\rm TE_{10}$ modes.

The linear characterization of the sample used for Fig. 4(c)-(d) in the main text is shown in Figure \ref{fig:s6}. Same as Fig. 3, the transmission spectrum and phase response are measured for the analysis of the quality factors. Comparing to Fig. 3, the BICs ($Q$-contrast) here is much weaker ($Q_{\rm i}$ value at regimes outside BICs is high), especially at the short wavelength region. This could be attributed to the reduced mode coupling among the modes. As a result, the parasitic mode competition cannot be efficiently suppressed, and the parametric oscillation can only reaches up to  14.6\% in conversion efficiency though the cavity modes are still in the strong over-coupling condition ($Q_{\rm i}/Q_{\rm c}>10$).

\section{h-OPO conversion efficiency analysis}

\begin{figure*}[h]
\centering
\includegraphics[width=1\linewidth]{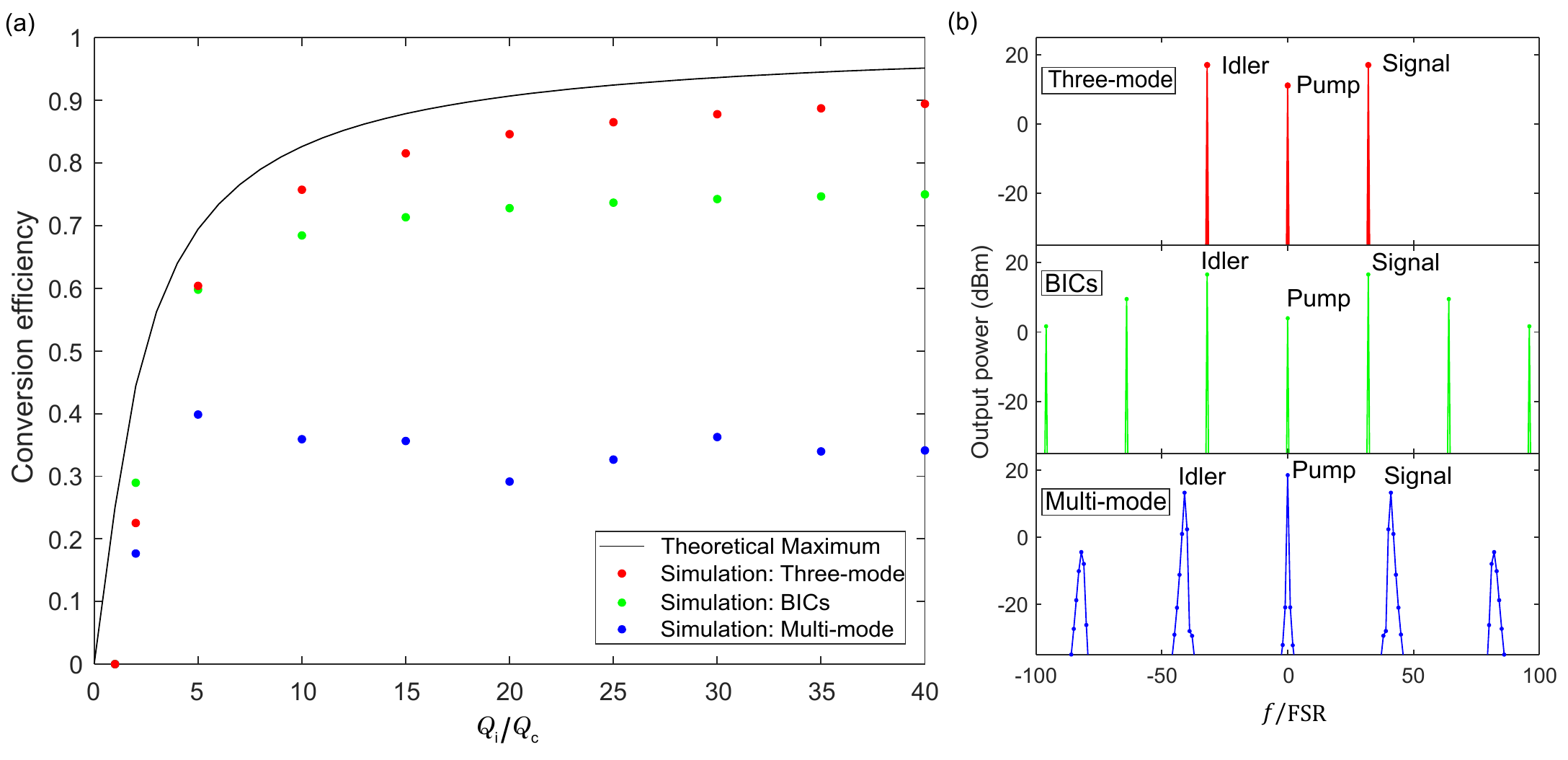}% Here is how to import EPS art
\caption{\label{fig:s7} Numerical simulation of conversion efficiency of h-OPO. (a) Simulated conversion efficiency as a function of quality factors. (b) Typical optical spectra corresponding to high conversion efficiency in three models. }
\end{figure*}

In this section, we numerically study the conversion efficiency in h-OPO. The maximal conversion efficiency in h-OPO is ultimately determined by the quality factors of microresonators \cite{sayson2019octave}:
\begin{equation}
    \eta_{\rm max}=(\frac{Q_{\rm i}/Q_{\rm c}}{1+Q_{\rm i}/Q_{\rm c}})^2,
     \tag{S.16}
\end{equation}
where the signal, idler and pump modes are assumed to feature the same $Q_{\rm i}$ and $Q_{\rm c}$. This theoretical limit is derived based on three-mode approximation, i.e., only signal, idler and pump modes are considered for interaction. To reach the theoretical maximum, proper parameters such as dispersion, detuning, pump power should be optimized in addition to quality factors.

By choosing: FSR=100.16 GHz, $Q_{\rm c}=0.8\times10^6$, group velocity dispersion $\rm \beta_2 =-46  ps^2/km$,  nonlinear coefficient of microring waveguide $\rm \gamma=0.76 W^{-1}m^{-1}$, pump power $P_0=$ 21 dBm, and different $Q_{\rm i}$ for active modes (see below), we performed a series of numerical simulations of the conversion efficiency in h-OPO, see Figure \ref{fig:s7}. The conversion efficiency in the plot represents the maximum value regarding to the pump-cavity detuning. 

The simulations were performed with an Ikeda map of a Kerr microresonator-waveguide system, where 512 cavity modes were considered. To perform three-mode simulation, we inactivate all modes (by setting their $Q_{\rm i}$ to be 0) except for mode -32, 0, 32 (counted from pump mode).  As a result, only the three modes can be effectively excited and the conversion efficiency can be achieved to be close to the theoretical maximum. 

To simulate our microresonator with BICs, the modes with number $n\times$32 are turned to be active ($n=\pm 1,2,3...$). In this case, high-efficient coherent h-OPO can still be generated but slightly (typically few percent) lower than the three-mode case, which can be explained as partial power is lost into high-order sidebands through four-wave mixing process. 

If 512 cavity modes are activated, i.e., multi-mode case, the maximum conversion efficiency is lower, which is not mainly limited by quality factors but by mode competition, see Figure \ref{fig:s7}(b).

These analyses demonstrate that the use of BICs can get the conversion efficiency in realistic multi-mode systems closer to the ideal situation of a pure three-wave h-OPO.

%\nocite{*}
%\bibliographystyle{apsrev4-2}
\bibliography{ref}% Produces the bibliography via BibTeX.

\end{document}